\documentclass[]{aastex631}

\usepackage [english]{babel}
\usepackage [autostyle, english = american]{csquotes}
\usepackage{wrapfig}
\shorttitle{Polstar ISM Science}
\shortauthors{Andersson et al.}

\begin{document}

\title{Ultraviolet Spectropolarimetry with Polstar: Interstellar Medium Science}

\author[0000-0001-6717-0686]{B-G Andersson}
\affiliation{SOFIA Science Center, USRA, NASA Ames Research Center, M.S. N232-12, Moffett Field, CA 94035, USA}

\author[0000-0002-0141-7436]{Geoffrey C. Clayton}
\affiliation{Dept. of Physics and Astronomy, Louisiana State University, Baton Rouge, LA 70803, USA}

\author[0000-0002-9829-0346]{Kirstin D. Doney}
\affiliation{Lockheed Martin Advanced Technology Center, 3251 Hanover Street, Palo Alto, CA, USA 94304}

\author[0000-0003-2017-0982]{Thiem Hoang}
\affiliation{Korean Astronomy and Space Science Institute, Daejeon 34055, Korea, and Korea University of Science and
Technology, Daejeon, 34113, Korea}

\author[0000-0002-1580-0583]{Antonio Mario Magalhaes}
\affiliation{Depto. de Astronomia, IAG, Universidade de S\~ao Paulo,
Rua do Matão 1226, S\~ao Paulo, SP 05508-090, Brazil}

\author[0000-0001-7482-5759]{Georgia V. Panopoulou}
\affiliation{California Institute of Technology, MC 350-17, Pasadena, CA, 91125, USA}

\author[0000-0003-2560-8066]{Huirong Yan}
\affil{Deutsches Elektronen-Synchrotron DESY, Platanenallee 6, D-15738 Zeuthen, Germany}
\affil{Institut f{\"u}r Physik und Astronomie, Universit{\"a}t Potsdam, Haus 28, Karl-Liebknecht-Str. 24\/25, D-14476 Potsdam, Germany}

\author[0000-0002-9071-6744]{Paul A. Scowen}
\affiliation{Code 667, Exoplanets \& Stellar Astrophysics Lab., NASA Goddard Space Flight Center, Greenbelt, MD 20771, USA}



\begin{abstract}

Continuum polarization over the UV-to-microwave range is due to dichroic extinction (or emission) by asymmetric, aligned dust grains.  Because of both grain alignment and scattering physics, the wavelength dependence of the polarization, generally, traces the size of the aligned grains.   Ultraviolet (UV) polarimetry therefore provides a unique probe of the smallest dust grains (diameter$<0.09\mu$m), their mineralogy and interaction with the environment. However, the current observational status of interstellar UV polarization is very poor with less than 30 lines of sight probed.   With the modern, quantitative and well-tested, theory of interstellar grain alignment now available, we have the opportunity to advance the understanding of the interstellar medium by executing a systematic study of the UV polarization in the ISM of the Milky Way and near-by galaxies.  The Polstar mission will provide the sensitivity and observing time needed to carry out such a program, addressing questions of dust composition as a function of size and location, radiation- and magnetic-field characteristics as well as unveiling the carrier of the 2175\AA\ extinction feature.   In addition, using high-resolution UV line spectroscopy Polstar will search for and probe the alignment of, and polarization from, aligned atoms and ions - so called ``Ground State Alignment", a potentially powerful new probe of magnetic fields in the diffuse ISM.

\end{abstract}



\section{Introduction} \label{sec:intro}

\begin{figure}
    \centering
    \plotone{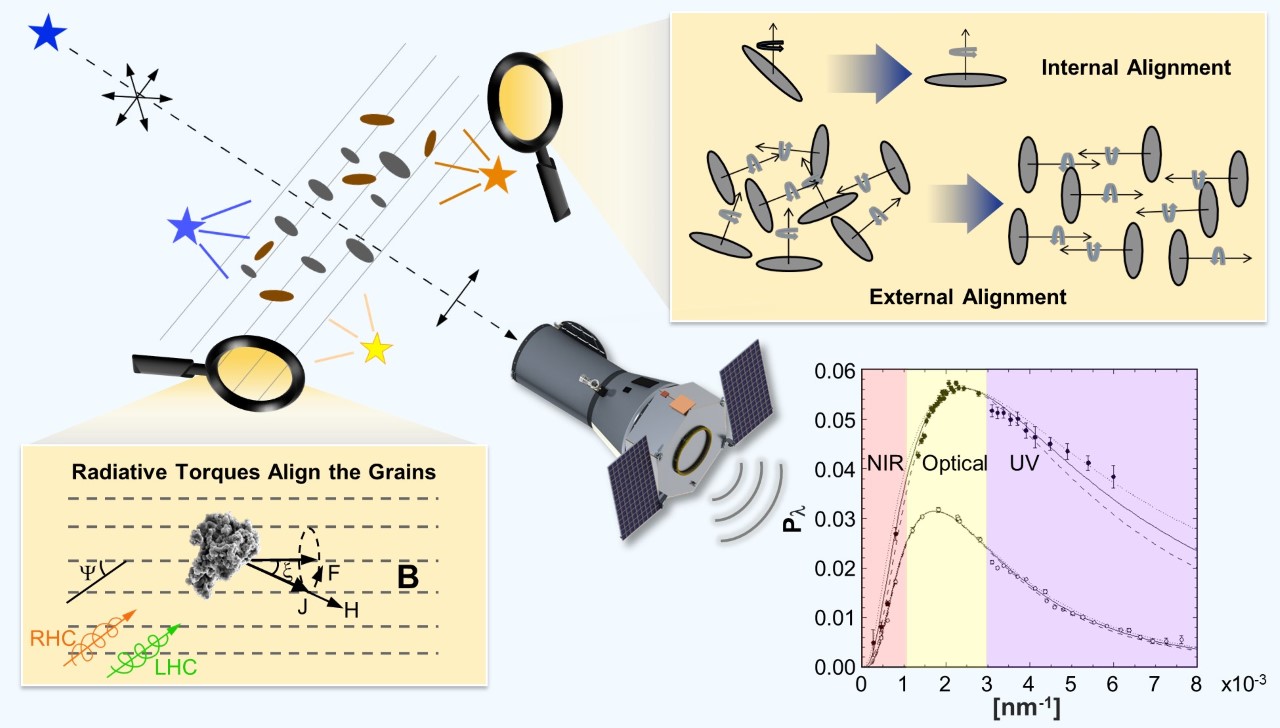}
    \caption{Interstellar polarization is due to elongated dust grains aligned [usually] with the magnetic field.  To cause polarization the grain’s spin axis must both be aligned with one of its symmetry axes (internal alignment) and with an external reference direction.  Radiative torques spin the grains up, and if the grain is paramagnetic - such as Silicates - it will experience both alignments via the Barnett effect (gray ovals upper left).  Carbon solids are, however, diamagnetic and are not expected to align under most circumstances (brown ovals).  The convolution of grain sizes, mineralogy and illuminating SED yields the observed polarization curve.}
    \label{fig:my_label}
\end{figure}

The interstellar medium (ISM) is permeated by magnetic fields and radiation, which interact with ubiquitous interstellar dust particles, of varying size, shape and mineralogy. The microphysical interaction of these components results in the macroscopic effect of dust-induced polarization of starlight. Since its discovery at optical wavelengths \citep{hiltner1949a,hiltner1949b,hall1949}, interstellar polarization has been known to arise from dichroic extinction by elongated dust grains aligned with the ambient magnetic field (see Fig. \ref{fig:my_label}). Interstellar polarization offers a unique method to probe the ISM magnetic field and study the intrinsic properties of grains (e.g. composition, size distribution). 

Interstellar magnetic fields not only provide the reference direction for most ISM grain alignment, and hence polarization, but are of direct interest, as they are expected to play a crucial role in the dynamics of the ISM, by providing additional pressure regulating star formation, as well as restricting and guiding gas flows \citep{HennebelleInutsuka}. In interfaces between hot and cool gas the magnetic field acts as an insulating layer preserving the neutral gas in hot environments \citep[e.g.][]{fox2004,fox2006,bga2004}.  Mapping of the polarization over a coherent area allows the field strength to be derived using the Davis-Chandrasekhar-Fermi method \citep{davis1951b,chandrasekhar1953}.  For UV polarization this will often focus on the diffuse or ionized gas (Sec. \ref{sec:SuSeP}).

The microphysics of how grains become aligned with the magnetic field has been the subject of study ever since the first detection of interstellar polarization.  A quantitative theory for grain alignment was proposed early on by \cite{davis1951a} (DG) based on paramagnetic relaxation in a rotating grain.  In the original formulation the grains were spun up by gas-grain collisions to rotation speeds corresponding to the thermal energy of the gas.  Based on cosmic elemental abundances we'd expect (which is, generally, also borne out by observations) that interstellar dust consists primarily of silicates and carbonaceous grains.  As noted by \citet{jones1967} the time scale for the paramagnetic relaxation for ``astronomical silicates" is too long (relative to the randomization by gas-grain collisions) to allow efficient alignment.  They proposed that the inclusion of sub-grains with higher magnetic susceptibility could speed up the alignment sufficiently to allow the observed polarization.  \citet{mathis1986} used this hypothesis to explain the sharp cut-off in small aligned grains required by the inversion of the polarization curve.   Because DG alignment is nominally more efficient for small grains \citep{draine2011} some mechanism is needed to stop grains smaller than about effective radius \textit{a}$\sim$0.05$\mu$m from being aligned. \citet{mathis1986} argued that this came about because the cut-off size is where the first ``super-paramagnetic inclusion" is statistically incorporated into a grain.   Indirect, empirical, support for this idea comes from the microscopy analysis of interplanetary dust particles \citep{goodman1995b}.  As an alternative (or complementary) to ``super-paramagnetic grains" \citet{purcell1979} proposed that if the grains could be spun up to rotation speed significantly above the thermal energy of the gas, thermal gas-grain collisions would not be as efficient in randomizing the grain orientation.  He suggested that the most likely source of such ``suprathermal spins" would be photoelectric emission from the grains and, especially, the ejection of newly formed hydrogen molecules \citep{vandehulst1948} from the grain surface.  

An important conclusion from the \citet{jones1967} work was that, based on very fundamental thermo-dynamical arguments, for DG alignment to work, the gas and the dust temperatures have to differ.  Over the following decades, observations \citep[e.g.,][]{jones1984,hough2008} showed alignment at high opacities where this condition is unlikely to be met.  Also, \citet{voshchinnikov2012} showed no correlation between the fraction of iron in the grains in a likely ferromagnetic state and the fractional polarization.  Theoretically, \citet{lazarian1999} showed that for grains with a finite internal temperature the internal excitation (e.g., phonons) will couple to the macroscopic dynamic of the grain, such that rather than achieving a suprathermal spin rate over a continuous period, the grain will flip frequently, reversing the direction of the ``Purcell rockets", and leaving the grain in a state of ``thermal trapping" at low angular momentum.

In response to these issues, a new alignment paradigm was developed based on the recognition by \citet{dolginov1976} that grains with a net helicity will be spun up by the difference in scattering of the right- and left-hand circular polarized light in an anisotropic radiation field.  Numerical modelling \citep{draine1996,draine1997} showed that radiative torques could spin the grains up to rotational speeds significantly above the thermal energy of the gas (``supra-thermal rotation").  Laboratory measurements \citep{abbas2006} empirically confirmed the mechanism.  \citet{lazarian2007} developed an analytical formulation of the theory able to reproduce the numerical results and provide a general, predictive framework for exploring and testing the mechanism.  A large number of specific tests of the theory were formulated and conducted over the following decade, all supporting the predictions of this radiative alignment torque (RAT) paradigm.  For a summary of the first $\sim$20 years of RAT alignment research see \citet{bga2015b}.  

The theory (and its observational testing) has continued to develop over the last decade with several extensions and clarifications, including the possibility of alignment along the radiation field direction \citep[a.k.a. ``k-RAT"][]{lazarian2019}, rotational disruption of dust grains from fast rotation rates \citep{hoang2018}, the influence of grain mineralogy on the alignment \citep[Andersson et al., 2021 (in preparation)][]{lazarian2020} and the possibility of a new mechanical alignment mechanism akin to the RAT process \citep{lazarian2021}.

\subsection{Radiative Alignment Torque (RAT) Theory}
\label{sec:RAT}

While complex in its detail, RAT alignment is conceptually simple:  A dust grain with a net helicity will be spun up if exposed to an anisotropic radiation field with a wavelength less than the grain radius.  The wavelength dependence comes about from scattering physics and can be seen in the standard Mie scattering plots (e.g. \citet{whittet2003}, Figure 3.1).  We note that while the efficiency of the torque transfer does not fully vanish at $\lambda>d$ (where d is the grain diameter), it falls of very rapidly, such that for a qualitative discussion the RAT alignment condition:
\begin{equation}
    \lambda<2a
\end{equation}
\noindent
where \textit{a} is the effective grain radius, is accurately fulfilled.

For a grain made of paramagnetic materials, the resulting rotation, via the Barnett effect \citep{purcell1979} will magnetize the grain.  The Barnett effect \citep[whose inverse is the Einstein - de Haas effect][]{richardson1908,einstein1915} is the mechanism by which a grain bulk with unpaired quantum mechanical spins (i.e. a \textit{paramagneic} material) will redistribute the total angular momentum of the system to achieve a minimal energy state. To do so, some of the rotation angular momentum is redistributed to spin-flips, leading to a net non-zero number of spins in the direction of the angular momentum, thus inducing a magnetization of the material.  This magnetization, in turn, will cause the grain angular momentum to Larmor precess around an external magnetic field.  The continued RAT actions over the Larmor precession will then align the grain angular momentum vector with the magnetic field.

To achieve significant polarization from an ensemble of dust grains two related, but separate, kinds of grain alignment must occur.  The alignment of the grain angular momentum vector with an external reference direction (above, the magnetic field direction) is referred to as ``external alignment".   In addition, ``internal alignment" of the grains is also required.  This signifies the alignment of the grain's angular momentum with one if its symmetry axes and ensures that an individual grain shows a time-constant projected profile towards an observer.  For a system under angular momentum conservation it is easy to show \citep[e.g.][]{purcell1979} that the minimal energy state of a rotating solid corresponds to rotation along the axis of maximum inertia (i.e. the minor axis).  Given an efficient internal energy dissipation mechanism in the grain, we can expect this state to be rapidly achieved. As shown by \citet{purcell1979} Barnett relaxation in a nutating paramagnetic grain is likely to be such an efficient energy dissipation mechanism.

From the above it is therefore clear that for RAT alignment to be efficient (and also required for DG alignment), the grains bulk must be, at least, paramagnetic.  Of the two main components of interstellar dust - silicates and carbonaceous materials - the former is well known to be paramagnetic, while the latter are expected to be diamagnetic \citep{heremans1994, wang2015} and not expected to experience either the Barnett effect or respond to externally applied magnetic fields.  If dust is efficiently reprocessed in the ISM composite grains are also possible \citep{hensley2020}, but will tend to be larger.

Hence, while silicate grains are expected to efficiently align and give rise to significant polarization, carbon dust should not.  Through second-order effects \citep{hoang2009a,lazarian2020,bga2021a} inefficient alignment of carbonaceous grains in high-radiation field environments is possible and has likely been observed in the carbon-rich circumstellar envelope of the asymptotic giant branch star IRC+10\ 216.

RAT alignment therefore provides a number of diagnostics of the dust grains in the ISM, their environment and ultimately the ISM magnetic field.  Because of the RAT alignment condition, the size distribution of \textit{aligned} grains is produced by a convolution of the underlying total size distribution of the dust \citep[which generally follows a steep power-law form][]{mathis1977} with the spectral energy distribution (SED) of the aligning radiation field.  While in the general, neutral ISM the shortest wavelength of light is given by the Lyman limit at 912\AA, corresponding to the ionization threshold of neutral hydrogen, for fully ionized gas shorter wavelengths of light can propagate and therefore paramagnetic grains with $a<0.045\, \mu m$ may be radiatively aligned.  Therefore, ultraviolet polarimetry uniquely probes the composition and abundance of the smallest grains (\textit{a}$<0.045\mu$m) and the hard (Extreme Ultraviolet; EUV, $\lambda<912$\AA) radiation field.

\subsection{The need for UV polarimetry to tackle open questions in ISM research}

The current state-of-the-art in space UV polarimetry was set by the combination of two instruments: 1) the Wisconsin Ultraviolet Photo-Polarimeter Experiment (WUPPE) which flew on the Space Shuttle as part of the Astro-1 and -2 missions \citep[][]{Nordsieck_1994} and 2) the Faint Object Spectrograph (FOS) aboard the Hubble Space Telescope (HST; e.g. \citet{allen1982,clayton1995}). A total of 28 lines of sight were observed in UV polarization \citep{martin1999}. The survey results revealed fundamental questions about the nature of interstellar polarization (Sec. \ref{sec:SuSeP}, \ref{sec:2175}). These questions remain unsolved to the present day. The development of RAT theory (Sec. \ref{sec:RAT}), combined with the outstanding progress in tracing magnetic fields in the ISM \citep{pattle2021} have added an array of open problems in ISM research: from constraining dust and environmental parameters of the ISM, to selectively tracing the magnetic fields in fully ionized gas.
We identify pressing challenges that require the next generation space UV polarimetry mission to be addressed:  
\begin{itemize}
    \item \textit{Determining how the alignment of the smallest grains varies with environment.} Because of the very limited sample of ISM lines of sight observed so far in the space UV  \citep[28;][]{martin1999}, and the heavy bias of that sample to the brightest stars (rather than well-defined ISM environments), the variations of the UV polarization curve with environment are poorly understood.  As noted above, grain mineralogy as well as radiation field environment will affect the polarization characteristics as well as variations in the overall grain size distribution.  As shown by \citet{Hoang:2019NAT} grains exposed to a strong radiation field may be radiatively spun up to high rotation speeds such that the centrifugal stress exceeds the grain tensile strength, resulting in the rotational disruption of large grains into smaller fragments and the change of the grain size distribution. The latter effect depends sensitively on grain alignment efficiency, grain composition and structure \citep[cf., e.g.][]{draine2020}.  By surveying a significant sample of carefully selected ISM lines of sight a better understanding of the variations in the grain size and property distribution can be achieved.  This is particularly important in the low-metalicity environments of the Magellanic clouds for which almost no data exist on their UV polarization \citep{Clayton_1996}.
    
     \item \textit{Understanding how small grains can become `extremely' aligned.} The wavelength dependence of the polarization (polarization curve) from UV to NIR is a powerful probe of dust grain properties. For most lines of sight, the shape of this curve follows the well-established Serkowski law (Sect. \ref{sec:SuSeP}). However, the WUPPE mission has found an intriguing exception: the polarization curve in the UV exceeds the standard law for a small number of lines of sight \citep{Clayton_1992}. This implies that small grains are somehow more efficiently aligned in these regions than in the general ISM. Two scenarios have been proposed to explain the Super-Serkowski UV polarization (SuSeP): (a) DG alignment in regions with enhanced magnetic fields, and (b) enhanced RAT alignment due to radiation shortward of the Lyman limit. To test these hypotheses, we need UV polarimetry towards a large, carefully selected sample of lines of sight, where environmental parameters can be well characterized.
     
    \item \textit{Constraining the spectral energy distribution (SED) of dust-aligning radiation, especially short-ward of the Lyman limit.} If the SuSeP effect can be correlated with enhanced EUV radiation fields, then the combination of the extinction curve, from Stokes I data, and the polarization curve can (under RAT alignment) be inverted to constrain the EUV field responsible for grain alignment. With a census of the hot stars close to the dust, from Gaia and Hipporcos \citep[e.g.][]{dezeeuw1999}, in several regions, estimates of the EUV part of the SED for O and B stars can be derived.  As this EUV radiation does not efficiently penetrate the ISM, such observations will provide unique information on the high-energy spectra of the hot stars.
    
    \item \textit{Identifying the carrier of the UV bump at 2175\AA.} The carrier of the well-known 2175\AA\ extinction feature is, from scattering physics and abundance constraints, required to be small grains dominated by the elements: C, Mg, Si and Fe and most likely carbon-based \citep{whittet2003}.  As probed by WUPPE and HST/FOC  \citep{wolff1997,martin1999} this feature can be, but is only very rarely (2 out of 28 line of sight), polarized.  Given the rareness of the effect and the diamagnetic nature of carbon solids, a significantly increased sample of well characterized lines of sight is needed to clarify the nature of the carrier of the feature, its polarization state and its possible relationship to the poly-aromatic hydrocarbon macro-molecules observed prominently in the near and mid infrared \citep{blasberger2017}.
    
    \item \textit{Studying dust characteristics in low-metallicity environments.}  Because the grain alignment is sensitive to the mineralogy of the dust grains the polarization curve will reflect the metallicity of the dust, both absolutely and relatively (by wavelength), and can provide unique constraints in the dust chemistry.  While a small number of ISM lines of sight have been probed in near-by galaxies, only two lines of sight have been observed in UV polarimetry for the LMC and none in the lower metallicity SMC.  With the enhanced sensitivity of Polstar, the UV polarization curves toward several dozen stars in the LMC and about a dozen stars in the SMC can - with reasonable exposure times - be well characterized.   
    
    \item \textit{Studying the behavior of dust under extreme circumstances.} Because grain growth in the diffuse ISM is very slow \citep{draine2009} and the the radiative spin-up of grains is dependent on the radiation field strength and the grain sizes (as well as tensile strength), transient radiation field enhancements such as near novae or supernovae may provide a mechanism to probe, both the tensile strength of the grains and the local (to the transient source) grain abundance and size distribution.  For instance, by observing a varying (UV) polarization curve of a [super]nova explosion, the host galaxy extinction of the event can be separated from the Galactic extinction. 

    \item \textit{Clarifying the origin of high polarization efficiency.} Whereas the last several decades have shown that the RAT paradigm is the dominant explanation for the observed ISM polarization, the absolute levels of the polarization remain an open problem.  As shown by \citet{planck2020,panopoulou2019}, the maximum polarization per unit extinction can significantly exceed the levels previously seen and predicted.  The level of $p_{max}$/A$_V$ presents a significant challenge to theory and may require, even for silicate grains, the inclusion of ferromagnetic sub-grains.  However, since the bulk of the grains - because of abundance constraints - must be silicate, with paramagnetic susceptibilities, there is likely to exist a threshold grain size below which no such ``super-paramagnetic inclusions" are available \citep[cf][]{mathis1986}.  Because UV polarimetry selectively probes the smallest grains, the hypothesis that the very high polarization fraction observed is due to super-paramagnetic inclusions can be directly tested by comparing optical to UV polarization in these regions.
 
    \item \textit{Testing ground state alignment as a probe for ISM magnetism.} As has long been established in laboratory laser-physics, atoms (and ions) with angular momentum $\geq 1$ (fine and/or hyperfine structure) can be aligned toward the magnetic field through resonant line scattering with anisotropic radiation field \citep[e.g.][]{happer1972}.  In consequence, absorption and scattering from such aligned atoms are naturally polarized arising from the difference in $\sigma$ and $\pi$ transitions, and the quantitative predictions of the resulting polarizations \citep{YLfine, YLHanle,yan2012} from various atomic lines demonstrated the diagnostic power of ``Ground state alignment" (GSA) in probing the geometry and strength of the magnetic field, especially in diffuse, low field strength, environments. Many of the best spectral lines for probing ISM magnetic fields with this GSA technique are located at UV wavelengths. UV spectropolarimetry therefore stands a unique chance to trace magnetic fields through the GSA technique. An obvious advantage is the location information immediately achievable with the spectrometry, making the magnetic tomography possible.

\end{itemize}
 
 \begin{figure}[ht]
     \epsscale{0.99}
     \centering
     \plotone{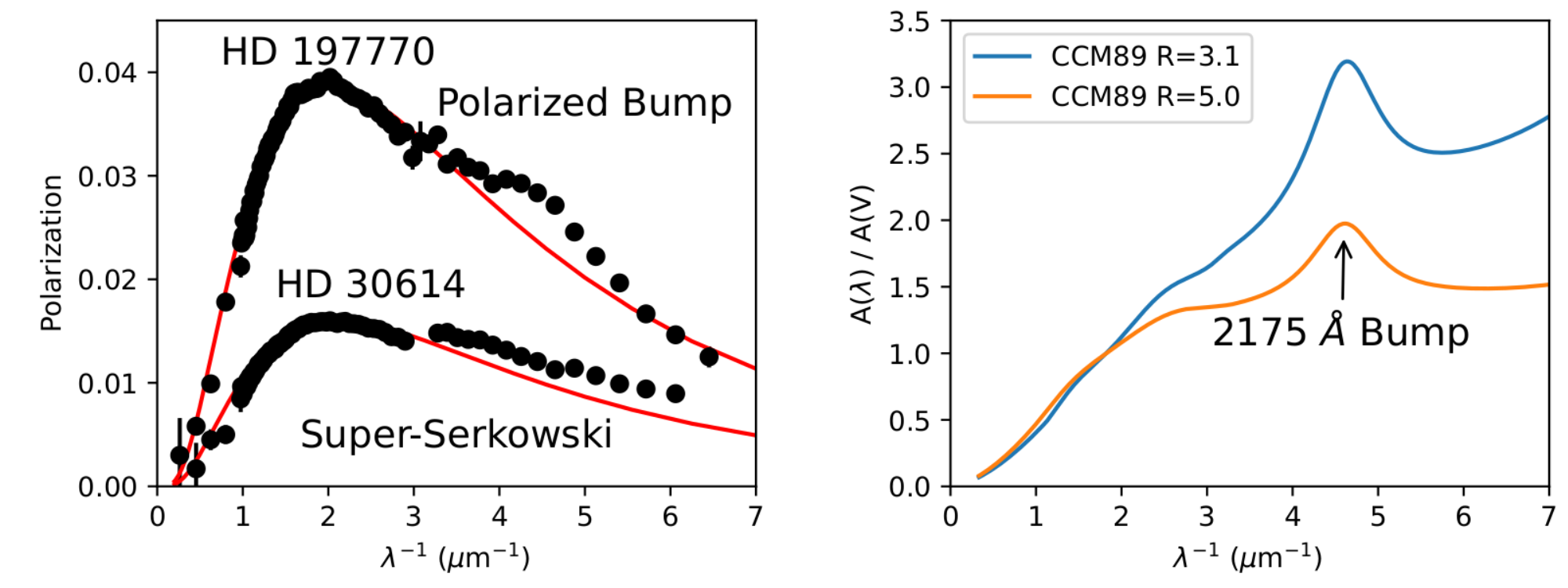}
     \caption{UV polarimetry will address many aspects of the nature of small dust grains and their interaction with the environment, including the Super-Serkowski UV polarization effect, the carrier of the 2175\AA \, extinction feature, and through a combination with the extinction curve (variations) the mineralogy of the grains as a function of size. \textit{Left:} an example of Super-Serkowski polarization, with data from a combination of WUPPE and optical measurements of the star HD30614 \citep[adapted from][]{clayton1995}. Also shown is an example of a sightline showing polarization of the 2175\AA \, feature \citep[data of the star HD197770 from][]{clayton1995}. \textit{Right}: Extinction curve from optical to UV showing variation for different values of $R_V$ \citep[from][]{CCM89}. }
     \label{fig:SuSeP_2175}
 \end{figure}
 
These problems can uniquely be addressed by polarization measurements in the space UV, often complemented by ground based optical observations.  Technical requirements for the dust-induced polarization include low to medium spectral resolution ($\sim$30-300) spectro-polarimetry -- the range between $\sim$1800-3200 \AA\ is particularly critical.  Because of the very limited existing sample of ISM UV polarimetry, efficient and flexible targeting and observations is needed. An unbiased sampling of ISM polarimetry over a broad range of environmental conditions will require a significant number of background stars to be observed at a very large range in intrinsic brightness, from near-by star forming regions to the faint stars in the small Magellanic cloud.  For the GSA polarization high spectral resolution (R$\geq$ 30,000) at high sensitivity is needed. 
For all science cases, excellent control of instrumental systematics (at a level of $\sim 0.1\%$ in polarization) is necessary. The Polstar mission \citep{Scowen2021} has been designed to provide the observing capabilities to accomplish this broad set of interstellar medium science objectives. 


\section{Primary ISM Science Cases for Polstar} \label{sec:scicases}

Ultraviolet polarization can probe a number of variables, of the dust, the environment - including radiation and magnetic fields - and the mechanisms for aligning the grains, or atoms (See sec. \ref{sec:GSA}). Below we describe our identified ISM science priorities for the proposed mission.  We stress that - in addition and because of the very limited, existing, UV polarization data - a wide discovery space for ISM astrophysics exists for a systematic survey of ISM UV polarimetry.

\subsection{Super-Serkowski UV Polarization} \label{sec:SuSeP}

The spectral shape of ISM polarization from the UV to the NIR shows a characteristic wavelength dependence which can be parametrized as
\begin{equation}\label{equ:serk}
p(\lambda)=p_{max} \cdot exp[-K \cdot ln^2(\lambda/\lambda_{max})]
\end{equation}
\noindent
\citep{serkowski1973}, where p$_{max}$ is the peak amount of polarization, occurring at the wavelength $\lambda_{max}$ and K - originally set to 1.15 - controls the width of the curve.  Subsequent work \citep[e.g.][]{codina1976,wilking1982} showed that the K-parameter varies over the ISM and that it is correlated \citep[e.g.][]{whittet1992} with $\lambda_{max}$.

As discussed by \citet{kim1995,clayton2003} the interstellar polarization curve can be derived from scattering theory, a grain-size dependent alignment efficiency and an underlying total grain size distribution.  Mie scattering \citep{mie1908} and its numerical refinements \citep[e.g.][]{herranen2019} are well established and understood.  The underlying total grain size distribution is also, generally, well established $-$ especially in the diffuse ISM $-$ as a power-law with a large negative exponent around -3.5 \citep{mathis1977}. Inverting the observed polarization curve, and comparing that to the inversion of the extinction curve \citep[e.g.][]{kim1995} it is clear that only relatively large grains are aligned. As paramagnetic relaxation theory predicts that smaller grains should be more efficiently aligned \citep{Draine_2011} the required cut-off was proposed to be due to the fact that regular silicates have too small a magnetic susceptibility of align efficiently and that grains only become aligned once at least one ``super-paramagnetic inclusion" (iron compounds or similar) was included in the grain bulk \citep{mathis1986}.  

As shown by \citet{kim1995} the smallest aligned grains, also grow with visual extinction, which in DG alignment might be explained by increasing collisional disalignment as the material gets denser.  Under RAT theory the small-grain cut-off in the general ISM, at \textit{a}$\sim 0.045 \mu$m is explained by the RAT alignment condition ($\lambda<2a$) and the fact that for neutral material no radiation short-ward of the Lyman limit ($\lambda=912$\AA) can propagate.  The reddening of the interstellar radiation field into a cloud explains the modeled growth in the cut-off grain size for higher extinction lines of sight.  Detailed studies of the ISM polarization curve as a function of extinction for a number of local clouds \citep{whittet2001,bga2007} support these conclusions.

Therefore, a detailed understanding of the shape and variations of the polarization (``Serkowski") curve, especially together with information about the over-all dust size distribution and the interstellar environment provides rich information about the grain mineralogy, the radiation field and gas characteristics \citep{vaillancourt2020}.

The Serkowski curve was initially, and is still mostly, based on ground-based optical observations.  With the advent of the WUPPE mission it has been extended into the UV for a limited number of lines of sight.  
As shown by \citet{clayton1995}, for most ISM sight-lines the UV polarization closely follows the extrapolation from the optical data.  However, for a subset of 7 lines of sight (out of the 28 probed by WUPPE and HST/FOS) the UV polarization significantly exceeds such an extrapolation from optical data.  This phenomenon is known as ``Super-Serkowski [UV] polarization" (henceforth ``SuSeP").

\begin{figure}
    \centering
    \plottwo{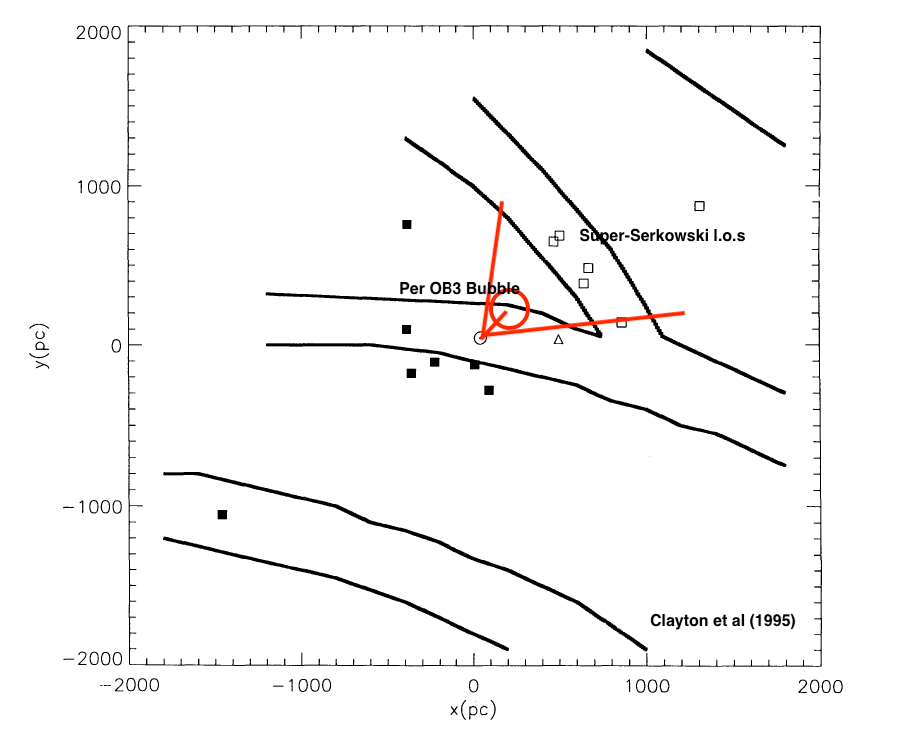}{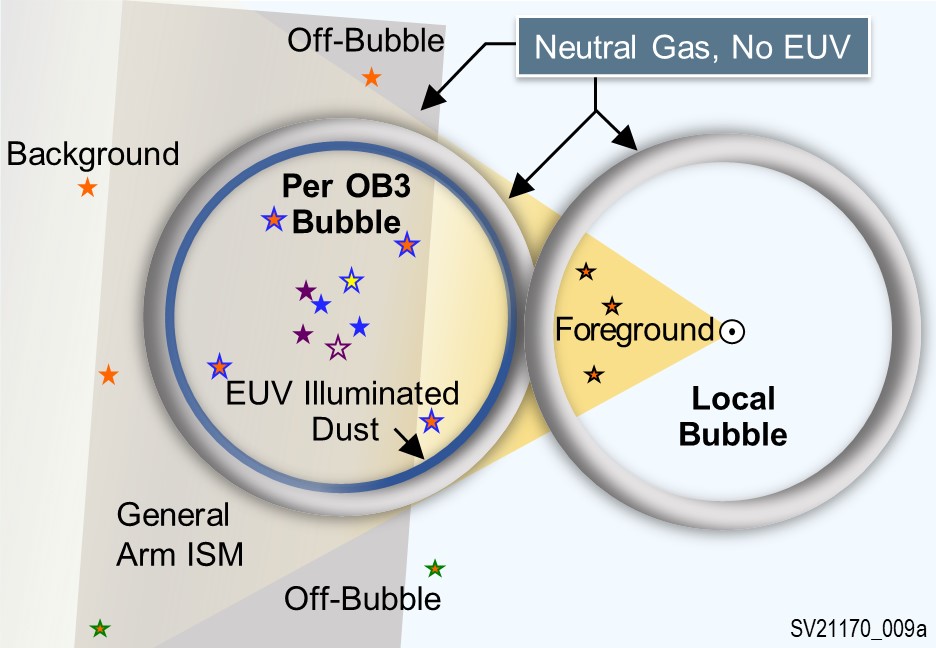}
    \caption{\textit{Left}: The SuSeP line of sight observed with WUPPE \citep{clayton1995} are located in the region of \textit{l}$\sim$95-150 and \textit{b}$\sim\pm$10. Overlaid on a schematic representation of the local Galactic disk the background stars (open squares) are found behind the Per OB3 super-bubble (red circle), as estimated from the observation by \citet{bhatt2000}. \textit{Right}: If Super-Serkowski UV polarization is due to very small grains aligned by EUV radiation we can understand the geometry of the WUPPE results with a simple model.  In fully ionized gas close to hot stars (such as the insides of OB-association bubbles) EUV light can align grains smaller than \textit{a}=0.045$\mu$m, enhancing the UV polarization.  Together with a census of the hot star in the OB-association (e..g. from Gaia data) the polarization curves, together with the extinction curves (Stokes I) can yield the EUV SEDs of the hottest stars, as well as probing the dust and magnetic field in the super-bubbles.}
    \label{fig:SuSeP}
\end{figure}

Two possible explanations have been put forward to explain the SuSeP effect.  \citet{hoang2014a} proposed that for regions of enhanced magnetic field strength, DG alignment of very small grains may become possible, which would enhance the UV polarization.  Another possibility, under RAT alignment, is that the aligning radiation field in some regions is extended below the Lyman limit and thus can align grains smaller than \textit{aers}$\sim 0.045 \mu$m.  As shown in \citet{clayton1995} almost all of the SuSeP lines of sight observed so far are located at \textit{l}=95-150$^\circ$ and \textit{b}=$\pm$10$^\circ$.  Interestingly, this region is delineated by a shell of neutral clouds surrounding the Per OB3 association \citet{bhat2011}.  Figure \ref{fig:SuSeP} (left panel) shows the extent of the Per OB3 super bubble on a top-down view of the local Galactic neighborhood including the location of the SuSeP lines of sight.  As this figure shows, all the SuSeP lines of sight pass through the Per OB3 supper bubble, which - if the interior of the bubble is fully ionized, and sufficient sub-Lyman flux is available from the OB association members, could explain the effect.  Figure \ref{fig:SuSeP} (right panel) shows a cartoon of this possible scenario.  \textit{Ab initio} RAT-based modeling (Lee 2021, private communication) shows that the presence of hot stars close to the lines of sight can reproduce a SuSeP-like polarization curve.  

Theoretical models of the evolution of super bubbles \citep{stil2009} indicate that the magnetic fields should be compressed and strengthened - in the direction perpendicular to the average field orientation -  in the wall of super bubbles.  Hence a combination of UV polarimetry and a census of the embedded hot stars are needed to fully differentiate between RAT alignment by EUV radiation and DG alignment in regions with enhanced magnetic field strengths. Further observations, including targeted observations of other near-by super bubbles, with varying OB association membership and ages, are needed to resolve the question of the origins of the SuSeP phenomenon.  PolStar is an ideal tool to perform such measurements.

As shown by \citet{clayton1995} the enhanced observed UV polarization is associated with small values of $\lambda_{max}$, a correlation that would be expected if the distribution of aligned grains is simply extended to smaller sizes.  Because the \citet{clayton1995} sample small and $\lambda_{max}$ is known to depend also on extinction and gas density \citep{whittet2001,bga2007,vaillancourt2020}, additional  data are needed to confirm this result.  We have carried out an optical polarization survey of $\sim$40 stars behind, inside, in front of, and to the sides of the Per OB3 bubble to indirectly test the origin of the SuSeP phenomenon.   As we show in Andersson et al., 2021 (in preparation) lines of sight probing the interior of the super bubble do \textbf{not} statistically show a smaller value of $\lambda_{max}$ than those probing only the foreground or stars to the side of the bubble.  Because of the general correlation of $\lambda_{max}$ with A$_V$, due to reddening into the cloud, larger samples of UV polarimetry for line of sight with full optical data are needed to fully characterize the SuSeP phenomenon and understand the degeneracies.  

If the explanation of the SuSeP effect is indeed the presence of sub-Lyman limit aligning radiation, then the effect probes the characteristics of the small grains (alignability, shape etc.) as well as selectively probing the magnetic fields in the ionized gas.  In addition, the effect would also provide a unique way to constrain the extreme ultraviolet (EUV) spectral energy distributions (SED) of the hot stars close to the observed polarization lines of sight.  Because EUV radiation is heavily obscured by the ISM, the stellar SEDs in this range cannot, usually, be probed from Earth \citep[see however][and references therein]{bowyer1995}.  Combining extinction curve measurements constraining the total grain size distribution (based on the Stokes I component of a polarization spectrum), with linear polarization spectra, quantitative RAT-based modeling can extract the SED of the aligning radiation.  Together with census of the hot stars in the associations \citep[e.g.][and equivalent Gaia surveys]{dezeeuw1999} accurate determinations of the EUV SEDs of the hot stars can be achieved.  It is important to note that because of the broad wavelength of Mie scattering, the polarization \textit{probing} this effect is \textit{not} at EUV wavelengths, albeit at space-UV wavelengths, accessible to PolStar.  

Because the observed polarization curve is a convolution of the size distribution of alignable grains with the aligning radiation field, UV polarization, in general (see also Sec. \ref{sec:SMC}), and SuSeP observations, specifically will also allow studies of the grain mineralogy and how it changes with environment - whether in the high-energy interior of local super bubbles or in the low metalicity environments of the LMC and SMC. Finally,  we note that SuSeP data selectively probes a limited part of the line of sight, either as a result of enhanced DG alignment or RAT alignment by localized EUV sources.  By observing several lines of sight probing each region showing SuSeP, the magnetic field in the that limited space can be constrained in 3D using the Davis-Chandrasekhar-Fermi method or its modifications \citep[e.g.][]{davis1951b,chandrasekhar1953,houde2009,CY16,Skalidis2021}.

A ground-based (optical) program to establish the nominal Serkowski-curve parameters for the target stars will be necessary to evaluate the effect and its causes in detail.  Because these observations can be done using broad-band photometric filters or low resolution spectro-polarimetry, only moderate sized telescopes are required for such surveys (cf. Andersson et al. 2021).
In order to quantify the amplitude of SuSeP, where present, the level of polarization for sighlines probing several OB association can be compared, as a function of wavelength, to the polarization spectra of stars probing regions where depolarizing effects are minimal and showing especially large fractional polarization \citep[][Sec. \ref{sec:p_max}]{panopoulou2019}.

\subsection{The 2175\AA\ Extinction Feature} \label{sec:2175}

Features in interstellar extinction curves give direct evidence of the dust grain compositions. 
The chemical composition
of dust grains has been a very difficult problem to
address since the available observational data do not provide strong constraints. Many solids have been suggested as constituents of the ISM dust
including ices, silicates, various carbon compounds, metals,
and complex organic molecules.
In the UV, there is only one spectral feature: the 2175 \AA\ extinction bump, discovered by \citet{1965ApJ...142.1683S}, and almost immediately ascribed to graphite grains \citep{1965ApJ...142.1681S}.
There have been no other discrete features found in the UV despite fairly sensitive search limits \citep{Clayton_2003}.
The bump feature is extremely strong and only a very abundant element such as carbon could be responsible. A strong case has been made for the “bump” near
to be due to sp$^2$-bonded carbon, as in graphite or PAHs \citep{Draine_2011} with some environmental variability \citep{whittet2004}, although other carriers cannot be conclusively ruled out \citep{draine1989,rouleau1997,ma2020}.

\begin{figure}
\begin{center}
 \includegraphics[width=0.99\textwidth]{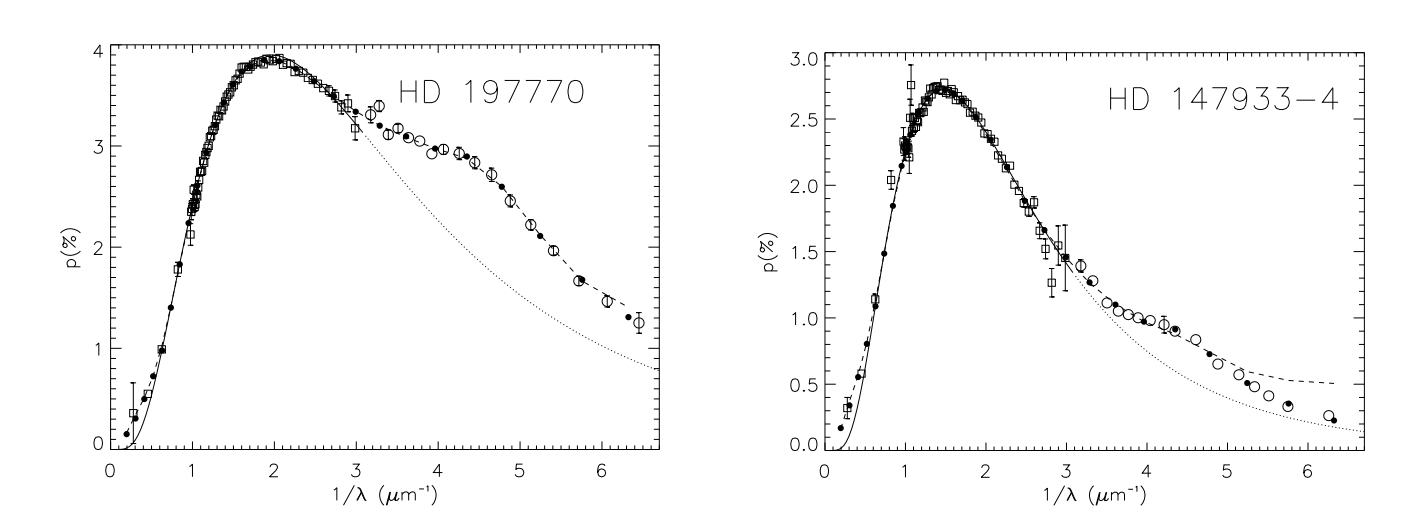}  
\end{center}
  \caption{Combination of WUPPE (open circles) and ground-based (squares) data for the stars HD 197770 (\textit{Left}) and HD 147933-4 (\textit{Right}). The solid line represents the
three-parameter Serkowski law fit to the visible and infrared data only, while the dotted line illustrates the extrapolation of this fit into the UV. The small filled circles are the interpolated data
used to determine the MEM model fit (dashed curve) \citep{wolff1997}.}
\label{fig:WUPPW_spectra}
\end{figure}

Some UV spectropolarimetry of ISM lines of sight is available from the {\it HST/FOS} and WUPPE. UV polarimetry was obtained for twenty-eight reddened lines of sight were observed with good S/N. Of these two, toward HD 197770 and HD 147933-4, showed a significant polarization excess through the region of the 2175 \AA\ bump \citep{wolff1997, martin1999}. The data are shown in Figure \ref{fig:WUPPW_spectra}.
The polarization excess seen for these two lines of sight 
is very small compared to the amount of
excess extinction present in the 2175 \AA\ bump, implying that the polarization efficiency for the bump grains is poor. A number of the other lines of
sight with UV polarimetry have high enough S/N to have detected polarization in the bump at the same level as seen in HD 197770 and HD 147933-4. Therefore, real variations in the alignment or grain shapes are present from one line of sight to another. 

The position angle does not change across the feature, i.e., it is the same as for the continuum polarization.This implies that somehow the 2175 \AA\ polarization feature is caused by the same population of grains responsible for the continuum polarization or there are two populations of grains with the same alignment direction. The continuum polarization is thought to be caused by silicate grains. It is possible to contrive a grain distribution that would produce the feature with only silicate grains but the second hypothesis of two similarly aligned grain populations is more likely. This is supported by the fact that the central position and width for Drude functions fitting the extinction bump and the polarization feature are identical within the uncertainties.
The polarization efficiency $\Delta$p/$\Delta \tau$ is at least 2 orders of magnitude
smaller than the theoretical maximum for perfectly aligned graphite carriers \citep{Martin_1995}, indicating that the polarization efficiency for these grains is very low.

Carbon is thought to be an important constituent of interstellar grains. It is highly depleted in the ISM. Graphite, amorphous carbon, and PAHs have been suggested as specific grain types. In particular, very small graphitic grains or PAHs have been suggested as the source of the 2175 \AA\ bump extinction. These grains would have to be aligned, albeit with low efficiency, to account for the observed polarization in the bump along two lines of sight. It has been suggested that small, aligned, oblate graphite grains could produce a polarization bump \citep{Draine_1988, wolff1997}. Polarization from PAHs has recently been claimed in the IR \citep{Zhang_2017}, albeit for the one line of sight probed, while the polarization line profile is well matched by the PAH feature, the position angle over the line rotates by a significant amount.  Further polarization studies of both the IR PAH lines and the 2175 \AA\ feature - preferably in the same regions are required to fully address the role of very small carbon grains in ISM polarization.


\subsection{Low Metallicity Environments} \label{sec:SMC}

It has become evident over the last few decades that the typical Milky Way R$_V$=3.1 extinction curve cannot be applied to dust extinction in other galaxies. The canonical extinction curve \citep[CCM][]{CCM89} was based on sightlines within 1 kpc of the Sun, while R$_V$ variations are observed out to 5 kpc \citep{Schalfly2017}; it is thus likely that the CCM curve does not hold throughout the Milky Way.
It is clear that CCM does not apply in our nearest neighbors, the Large and Small Magellanic Clouds (LMC and SMC) \citep{Gordon_2003}. Some lines of sight in the LMC are very CCM-like but the lines of sight near the giant star-forming region, 30 Dor, have weaker 2175 \AA\ bumps and steeper far-UV extinction. The extinction curves measured for SMC lines of sight are even more extreme. With the exception of one CCM-like sightline, the SMC dust extinction displays very steep far-UV extinction and weak or absent 2175 \AA\ bumps. The most likely explanation for the significant differences seen in the extinction characteristics in the Galaxy and the Magellanic Clouds is their global metallicity differences. Lower galactic metallicity means the abundance of metals available to make grains is less. The gas-to-dust ratio as measured by the amount of hydrogen is also higher in the LMC and SMC. 

A small survey of $\sim$20 lines of sight in the LMC found that the wavelength dependence of interstellar polarization in the optical (0.3-0.9 $\micron$) was consistent with the Serkowski relation for Galactic stars \citep{Clayton_1983}. Only two lines of sight in the LMC have been observed in the UV \citep{Clayton_1996}. Both lines of sight have small values of $\lambda_{max}$ which in the Galaxy are associated with super-Serkowski polarization curves. The uncertainties in the these observations is relatively high but both lines of sight are consistent with super-Serkowski behavior. 

Observations of the wavelength dependence of interstellar polarization toward a small sample of six stars in the SMC indicate that they fit the Serkowski relation with small values of $\lambda_{max}$ except for the sightline toward AZV 456, which is notable for having a CCM-like extinction curve \citep{Rodrigues_1997}. That sightline has a value of $\lambda_{max}$ more typical of Galactic lines of sight. There are no UV polarimetry data for the SMC.

With the sensitivity provided by Polstar significant samples of UV spectropolarimetry of lines of sight in the LMC and SMC will be possible, with - at least - several dozen in the LMC and a dozen star in the SMC.  These new observations will be key to determining how metallicity affects the nature of interstellar dust grains.

\subsection{Variations in absolute polarization efficiency} \label{sec:p_max}

The precise composition and physical properties of interstellar dust grains are highly uncertain, even for
the most well-studied population of grains existing in the diffuse ISM. Our understanding relies heavily on
forward-modeling the microphysics of dust to reproduce observables such as the interstellar extinction curve and the thermal dust emission SED in the FIR \citep{Draine2003}. 
Recent results from FIR and optical polarimetry revealed that both the level of polarization of dust emission and the ratio of polarized emission in the FIR to polarized extinction in the optical are inconsistent with previous expectations \citep{Ashton2018,planck2020}. New dust models are emerging that can accommodate these constraints, \citep{Guillet2018,draine2020}. The new models are informed by the Planck results and are constructed to reproduce all existing multi-wavelength constraints. The area where constraints for models are particularly lacking is the UV.

Polstar will allow for a systematic and precise characterization of the polarization curve in the UV. The efficiency of polarized absorption in the optical has been instrumental in calibrating new models. The maximum observed efficiency, defined as the ratio of $p_{max}/A_V$, can be linked to the level of alignment of the grain population that produces the polarization. The maximum value has recently been revised based on Planck data and optical measurements targeting lines of sight with well-constrained properties along the LOS \citep{planck2020,panopoulou2019}. By carefully selecting lines of sight where the optical polarization efficiency is maximum, and observing stars towards those lines of sight with Polstar, we will be able to provide a `gold-standard' UV-optical polarization curve. This will set the constraints that all models should adhere to. We will be able to measure if there is intrinsic variation in the UV/optical curve for these high-p-efficiency lines of sight, allowing us to determine how the alignment efficiency of the small grains is related to that of the larger grains - and whether this relation is consistent with new dust models. 

Polstar will enable studying variations of the polarization efficiency beyond the Milky Way as well. Currently, there are very few extragalactic lines of sight with observed high polarization efficiency. In cases such as M31 and NGC 3186, values of $p_{max}/E(B-V)$ much higher than the local 13\% were found \citep[][]{Leonard2002,clayton2004}. Although those specific targets are beyond the likely reach of Polstar, the proposed LMC and SMC surveys with Polstar will provide additional constraints on the polarization efficiency in low metallicity environments.

\subsection{Ground State Alignment}\label{sec:GSA}
 
\begin{figure}
 \includegraphics[width=0.45\textwidth]{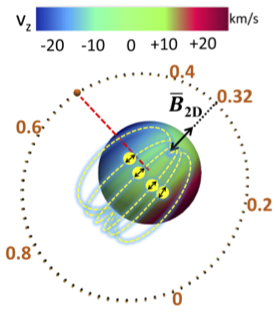}  
 \includegraphics[width=0.5\textwidth]{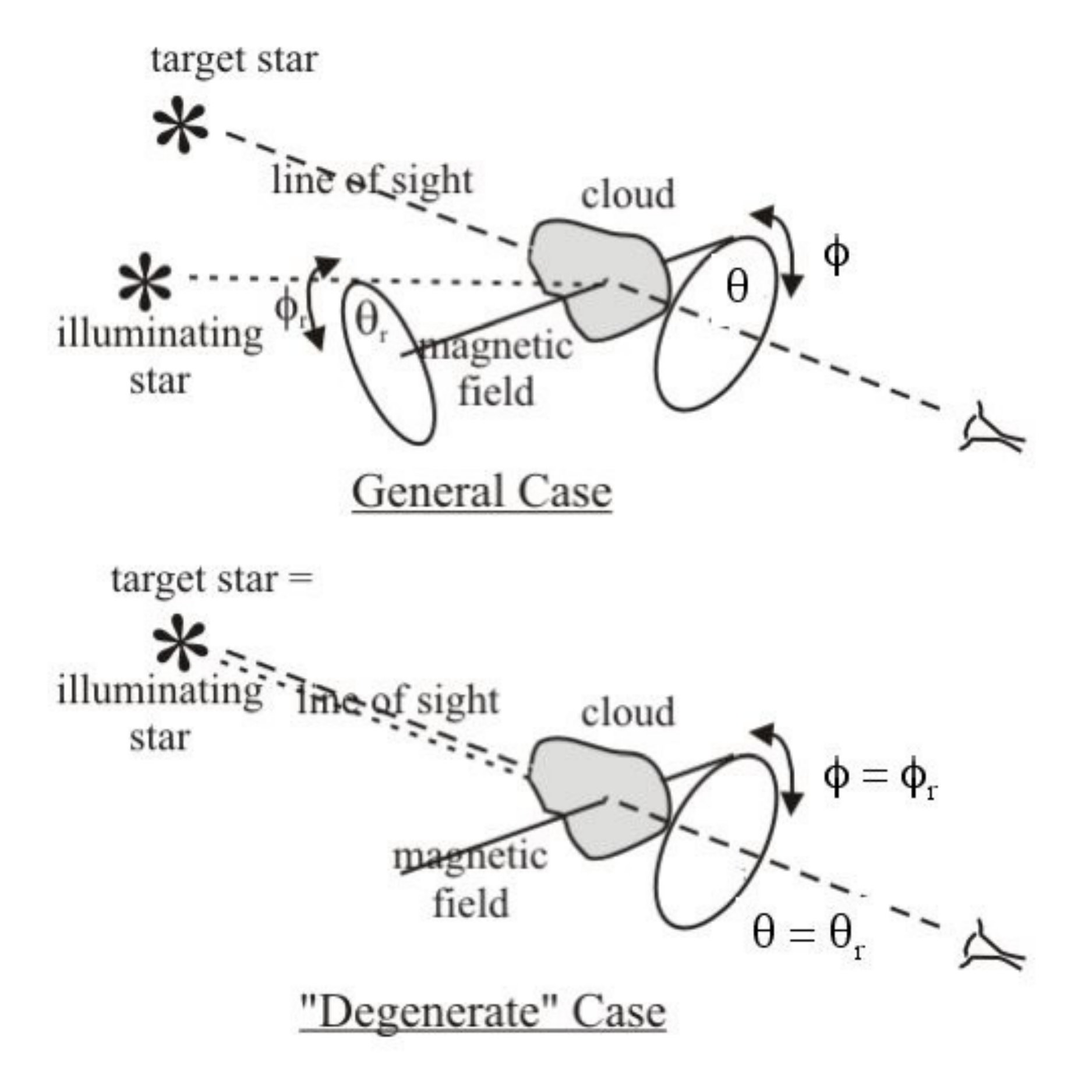}   
  \caption{\small \textit{Left}: 3D topology of magnetic field in $89~Her$ post-AGB binary system. The color scale indicates the line-of-sight velocity ($v_z$) of the medium. The inferred $3D$ magnetic field directions for different orbital phases are displayed \citep[from][]{GSA_obs20}. \textit{Right}: Typical astrophysical environment where GSA can happen. A pumping source deposits angular momentum to atoms and induces GSA. Observed polarization depends on both $\theta_r$ and $\theta$. In general, there are two situations: ({\it upper}), the alignment is produced by a pumping source while we observe another weak background source whose light passes through the aligned medium; ({\it lower}), the background source coincides with the pumping source, in this case, $\theta_r=\theta$.\citep[from][]{Yan19}}
\label{abspol_geom}
\end{figure}

GSA has been identified as an innovative way to determine the magnetic field in the diffuse medium. Atoms can become aligned in terms of their angular momentum and, as the life-time of the atoms/ions we deal with is long, the alignment
induced by anisotropic radiation is sensitive to weak magnetic
fields ($1G > B > 10^{-15}G$, \citealt{Yan_rev12}), a range encompassing the level of the magnetism in the ISM. The potential of GSA has been clearly revealed by the detection on a binary system where 3D magnetic fields are precisely mapped for the first time via polarization of absorption lines (see Fig.\ref{abspol_geom}a)\citep{GSA_obs20}.

Most of the resonance absorption lines are in the UV. A UV polarimeter with high spectral resolution (R $> 30,000$) will thus provide an incomparable opportunity for precision magnetic field measurement, which no other current instruments can offer. Particularly, the high spectral resolution allows simultaneous determination of both velocity and magnetic field, filling the gap of 3D magnetic tomography in ISM, which is so far missing \citep{Yan19}.

UV absorption lines are polarized through GSA exclusively. Any polarization, if detected, in absorption lines, would be a tell-tale indicator of alignment, which would necessarily trace the magnetic field, as no other mechanisms can induce polarization in absorption lines. 
If the pumping star is along the line of sight, as in the
central star of a reflection nebula, this is the so-called
``degenerate case" (see Fig.\ref{abspol_geom}b), where the
position angle of the polarization gives the 2D magnetic field in the plane of sky, and the degree gives the angle to the
line of sight. In the more general case, an observed cloud might be pumped
from the side, the positional angle of magnetic field is
available with $90^o$ degeneracy, while the derivation of the full magnetic
geometry requires measuring two lines either from the same species or
from two different species.  
With a high resolution spetropolarimeter, the 3D direction of the magnetic field along with the pumping radiation direction can be extrapolated by combining the polarization of two lines or the polarization of one line with their line intensity ratio  \citep[the latter is influenced by the magnetic field as well][]{YLfine, ZYR18}. 
With the knowledge of the degree of polarization and/or $\theta_r$, the angle between the magnetic field and line of sight, the 90 degree degeneracy can be removed.

The interstellar magnetic field is turbulent with velocity and magnetic fluctuations ranging from large injection scales to small dissipation scales. High resolution spectroscopy and spectropolarimetry combined bring forth a wealth of information on interstellar turbulence. Most magnetic diagnostics render only averaged mean magnetic field on large scales. In this respect GSA fits a unique niche as it reveals the small scale structure of the magnetic field. The statistical characterization of 3D magnetic turbulence can inform many problems including star formation, cosmic ray propagation and interstellar chemistry.


A possible detection of GSA in the UV would open up new potential for tracing local magnetic fields in a variety of environments where our understanding of the magnetic field geometry is limited. Examples include disks, PDR regions and the Local Bubble. One interesting case is circumstellar disks, for which grain alignment has been found unreliable \citep{Tazaki2017}. In the case of pre-main sequence stars, pumping conditions are similar to those for comets in the Solar System  \citep[see][]{Shangguan2013}: pumping rates on the order of 0.1 - 1 Hz, and realignment for fields greater than 10 -100 mGauss.  Conditions here are apparently conducive to substantial populations in CNO metastable levels above the ground term: \cite{Roberge:2002fk} find strong absorption in the FUV lines (1000 - 1500 \AA)
of OI (1D) and NI and SII (2D), apparently due to dissociation of
common ice molecules in these disks (these are also common in comet
comae).  Since these all have total angular momentum quantum number $>$1, they should be pumped, and
realigned.  This presents the exciting possibility of detecting the
magnetic geometry in circumstellar disks and monitoring them with time.

The magnetic realignment diagnostic can also be used in resonant and fluorescent scattering lines.  This is because the
alignment of the ground state is partially transferred to the upper
state in the absorption process \citep{YLhyf}.  In the cases where the direction of optical pumping is known, for example in planetary systems and circumstellar regions, magnetic realignment itself exhibits line polarization whose positional angle is neither perpendicular or parallel to the incident radiation. This deviation depends on the magnetic geometry and the scattering angle.  The degree of polarization also depends on these
two factors. In practice, GSA can be identified by comparing the polarizations from alignable and non-alignable species, which do not trace the magnetic field. There are a number of fluorescent lines in emission nebulae that are potential candidates \citep[see][]{Nordsieck:2008kx}. Reflection nebulae would be an ideal place to test the diagnostic,
since the lack of ionizing flux limits the number of levels being
pumped, and especially since common fluorescent atoms like NI and OI
would not be ionized, eliminating confusing recombination radiation.

GSA is usually not directly sensitive to the magnetic field strength. The exception from this rule is a special
case of pumping photo excitation rate being comparable with the Larmor frequency \citep[see][]{YLHanle,GSA_obs20}. However, GSA could be used to study the magnetic field strength as done for dust-induced polarization, via the DCF technique and its modifications (Pavaskar et al. 2021, in preparation). 

\subsection{Time Dependent effects}\label{sec:SN}

An exciting, but challenging, opportunity for UV spectro-polarimetry has opened up with the realization that strong radiative spin-up of dust grains may generate angular momenta, and centrifugal forces, high enough to shatter dust grains \citep{hoang2018}.  For transient sources, in particular, this can provide important insights both to dust dynamics and the intrinsic properties of the transient sources, such as novae and supernovae.

\begin{figure}
 \includegraphics[width=0.45\textwidth]{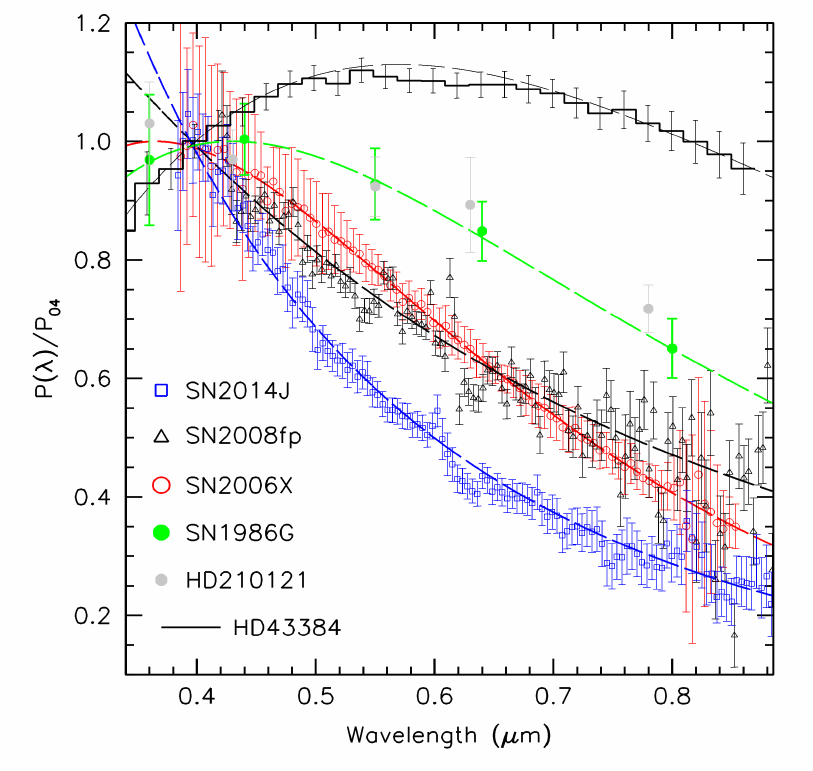}  
 \includegraphics[width=0.5\textwidth]{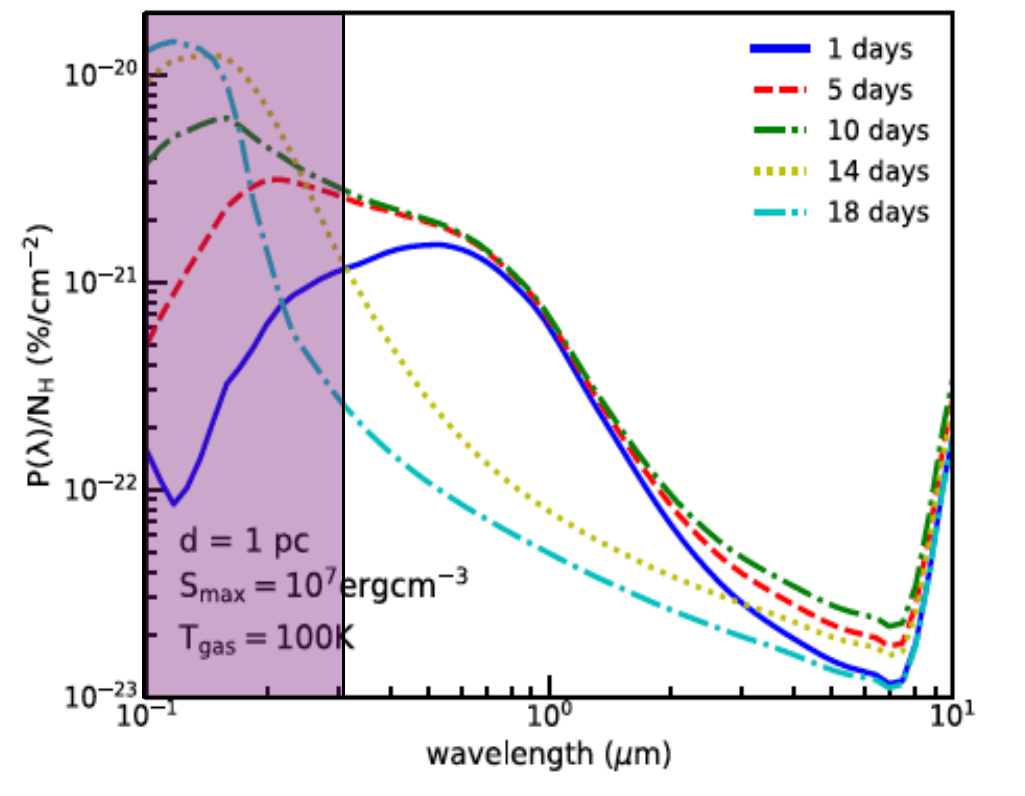}   
  \caption{\textit{Left}: Observations of recent extra-galactic supernovae often show polarization curves reminiscent of Serkowski curves, but with very small $\lambda_{max}$ values (e.g. SN1986G and SN2006X; reproduced with permission from \citet{patat2015}).  \textit{Right}: Numerical models of the predicted, time dependent, polarization curves under radiation-induced rotational disruption, where successively smaller grains are spun up to rotation speeds beyond their tensile strength  show a striking resemblance to the observed polarization curves.  In this example the distance to the supernova from the host-galaxy dust was set to 1 pc, with a gas temperature of 100 K and a dust grain tensile strength of 10$^7$ erg\,cm$^{-1}$. The purple area shows the wavelength range only accessible from space. Adapted, with permission, from \citet{giang2020}. }
\label{SNpol_fig}
\end{figure}

As shown by \citet{patat2015} the optical polarization curves towards extra-galactic supernovae (SNe) often show a very blue polarization, often rising beyond the atmospheric cut-off at $\sim$3200 \AA.  While some of this may be due to scattering, several of the spectra presented by \citet{patat2015} show indications of Serkowski-like shapes, but with very small values of $\lambda_{max}$ (Fig. \ref{SNpol_fig}).  \citet{giang2020} have shown that this may reflect a time evolution of the dust in the host-galaxy foreground, where over the first several weeks after the explosion radiative grain disruption ruptures increasingly small grains shifting the polarization curve increasingly to the blue.  As shown in Figure \ref{SNpol_fig} this evolution would rapidly move the peak of the polarization curve into the UV.

Monitoring of the polarization curves of bright supernovae (or Galactic novae) over the first few weeks, particularly in coordination with ground based polarimetry, promises to provide unique information on the host-galaxy extinction and dust characteristics.  If suitably bright SNe in (scheduling-wise) advantageous sky locations are available, Polstar would be able to uniquely contribute to such observations. Because of the uncertain availability of bright SNe during the prime mission duration of the Polstar mission, this possibility falls under the category "Enabled Science" and is not included as a design or science driver for the mission.

\section{Polstar ISM strategy} \label{sec:obs}

The orientation of the magnetic field with respect to the line-of-sight and with respect to the aligning radiation field can be probed with measurements of the linear polarization of specific atomic transitions between 122 - 200~nm, such as CI, CII, and SiII. Line-of-sight observations towards ten FUV-bright stars will test the theoretical expectation that ISM absorption lines seen in the stellar spectrum are the result of atoms in a magnetic region and will produce polarization from the GSA mechanism.

However, to understand the effect of polarization on the environment of the ISM requires spectroscopy of the continuum. By measuring the wavelength dependence of continuum linear polarization over 122 - 320 nm the minimum size and fraction of grain alignment, the shape, and optical properties of the aligned grains, and consequently the magnetic field orientation in a given environment can be determined. Observation of about 30 stars, probing each of five OB association super-bubbles: Perseus OB3, US, Upper Centaurus Lupus (UCL), Lower Centaurus Crux (LCC), and Ori OB1, will allow determination of how the alignment of the smallest interstellar grains occurs. In the Large Magellanic Cloud (LMC), Small Magellanic Cloud (SMC), and high-Galactic latitude material, measurements of the sized distribution and minimum size of aligned grains and the magnetic field orientation will shed light on how low metallicity affects dust grain alignment. 

A focus on measurements around 280~nm will provide an estimate of the excess of UV linear polarization relative to optical/NIR, and will test the SuSeP theory. Determination of the correlation between the excess UV polarization and proximity to a EUV source will accomplished through measurements of $\sim$12 stars in each of 10 nearby molecular clouds and star formation regions: S. Coalsack, Chamaeleon, rho Oph, Lupus, R CrA, Taurus, Perseus, OMC-1, and two reflection nebula regions. These sources are chosen as regions of maximal efficient dust emission, and observation of these sources can also inform on the mystery of the carrier of the 2175\AA\ bump. By measuring the column density and polarization of this feature as a function of extinction and PAH emission, new insight into the composition of the carrier and where and how it becomes aligned can be gained; this study requires correlated observations with archival observations from Spitzer in the IR to determine PAH emission and from HST and IUE in the optical to get the extinction curve of the unreddened sources.  Many of the unreddened comparison spectra, for the ``pair-method" for deriving extinction curves \citep{whittet2003}, will be achieved by observing the stellar Polstar targets also with the low-resolution ``Channel 2" spectrograph. 

To achieve the objectives described above requires single observation of the sources with a spectropolarimeter that can measure linear polarization from 122 - 320~nm with a resolving power of at least R = 30; and in the case of absorption line measurements linear polarization from 120 - 200~nm at an R $\geq$ 30,000. To accurately assess the polarization effect, a polarization precision of 0.1\% is required. The time on source is determined such that a maximum photometric signal-to-noise ratio (SNR) of about 1300 at 2175\AA\ for continuum observations and an SNR of 500 for atomic line observations is achieved. The Polstar instrument, a detailed description of which is given in Gilbertson \textit{et al.} 2021 (in preparation), will be able to meet these science needs. Polstar will enable measurement of all four Stokes parameters (I, Q, U, and V) using either a high-resolution FUV Channel (122 - 200 nm) or a low-resolution broadband Channel (180 - 320~nm). Moreover, at the planned orbit all anticipated sources are observable at least once throughout the year for long enough to meet the SNR required. 

\subsection{Synergy with Stellar Astrophysics Science Cases for Polstar}\label{Sec:decomp}

Polstar's broad wavelength coverage will enable a wide variety of investigations in addition to interstellar medium studies, in the physics of hot stars involving mass loss, the effect of binarity on winds, magnetospheres, and rapid rotation \citep[][]{Gayley2021,Peters2021,Louis2021,Jones2021,Shultz2021}, and in the properties of protoplanetary disks \citep{Wisniewski2021}.  Studying the same targets under a joint perspective will be advantageous for both stellar and interstellar programs.  On the one hand, measuring the ISM contribution is essential for understanding the intrinsic stellar polarization arising from different mechanisms. On the other, the ISM science benefits from the increased sample sizes, as many of the ISM effects are still at the parametric description stage where more details and a wide variety of environments will add to the observational constraints on the models.  Such synergies are, however, only viable if a clear separation of the stellar and interstellar polarizing effects can be achieved.  In addition to the dichroic extinction, responsible for the Serkowski curve, discussed above, the continuum effects likely to be involved in such a decomposition include Thompson scattering by free electrons in extended hot stellar atmospheres \citep[e.g.][]{chandrasekhar1946}, Rayleigh scattering by very small grains, either in circumstellar material \citep[e.g.][]{kastner1996} or in reflection nebulae \citep[e.g.][]{matsumura2011,bga2013}.  Line effects involve GSA polarization in the ISM (Sec \ref{sec:GSA}), Hanle and Zeeman effects in the stellar sources and broad, or quasi-lines, such as the 2175\AA\ feature (Sec. \ref{sec:2175}) and the line-blanketing effect in the stellar atmosphere \citep[e.g.][]{taylor1991}

Most of these effects have well known spectral dependencies including the Serkowski form  \citep[see eq. \ref{equ:serk} and generalizations proposed by][to include the Super-Serkowski effect]{martin1999}.   Thompson scattering is independent of wavelength while Rayleigh scattering displays a steeply rising polarization into the blue \citep[p$_R\propto \lambda^{-4}$; cf.][]{rybicki1979}.  A similarly well-constrained spectral shape applies for the 2175\AA\ feature.  
The majority of stellar polarization effects are spectral line effects with implied narrow wavelength ranges.  Even for quasi-continuum effects, such as line blanketing, the intrinsic wavelength dependence of the effect is nominally known.  The lines most likely to show strong GSA polarization are unlikely (due to ionization balances) to also originate in the atmospheres of hot stars.  In addition, we would generally expect that the position angles of the intrinsic stellar polarization and the polarization arising in the interstellar medium are unrelated.  

To decompose the different spectral components of the polarization we would, initially, assume that each component has a fixed position angle\footnote{For complex interstellar lines of sight, where the magnetic field orientation varies with distance and gas density \citep[cf.][]{hough1988}, more than one ISM term may be needed, although this is unlikely to apply to the low column density lines of sight observed in the UV.} and optimize the combination of Stokes parameters $q, u$: 
\begin{eqnarray}\label{equ:decomp}
q_{tot} = q_T + q_R(\lambda) + q_S(\lambda) + q_{line}(\lambda)\\
u_{tot} = u_T + u_R(\lambda) + u_S(\lambda) + u_{line}(\lambda)
\end{eqnarray}
where, the subscripts indicate Thompson (T) and Rayleigh (R) scattering, dichroic extinction (Serkowski; S) and various line polarization effects, against the observations.  The individual Stokes parameters are, as usual, given by 
\begin{eqnarray}
    q_i = p_i(\lambda) cos(2\theta _i)\\
    u_i = p_i(\lambda) sin(2\theta _i),
\end{eqnarray}
where $\theta_i$ is the position angle of each contributing mechanism.
The complexity of the applied function and its validity can be evaluated using F-tests \citep[e.g.][]{lupton1993} to quantitatively find the optimum solution.

More complex stellar polarization behaviors are possible: rapidly rotating stars that show a sharp increase in the polarization toward shorter wavelengths; axisymmetric models of winds, with possible position angle flips; hot star binaries - which in general will show chromatic effects even with electron scattering, because of the relative weightings by the two different SEDs. However, 
the general procedure outlined here is likely to be able to decompose the main components, given high-enough signal-to-noise data.  Deviation from flat continuum stellar polarization is discussed in the other Polstar white papers \citep{Peters2021,Gayley2021,stlouis2021,Jones2021,Shultz2021}.  At moderate signal to noise, these analyses may not provide complete or high-quality decomposition of the source terms.  They can still, at minimum, allow both ISM and stellar astronomers to flag the observations as having systematic effects and eliminate them from sample analyses where highly accurate determinations of either the ISM or stellar effects are required. 

Spectro-polarimetric decompositions of the kind discussed here have been successfully carried out for data combining stellar, circumstellar and interstellar polarization components in different combinations by several groups  \citep[e.g.][]{taylor1991,Andersson1997,Hoffman1998,nordsieck2001,lomax2017,Fullard2020}
We can therefore expect to use the broad wavelength coverage and high sensitivity of Polstar to decompose the stellar and interstellar contributions with confidence.  Specifically, Polstar will benefit from the extension of its spectral coverage (down to 1220\AA) compared to that of WUPPE (1500\AA) which will allow a clearer separation of the extended Serkowski curve from stellar effects, and also access to FUV stellar polarization due to gravity darkening of the disks of the many rapidly rotating stars among Polstar's stellar targets.

\begin{figure}
\begin{center}
 \includegraphics[width=0.5\textwidth]{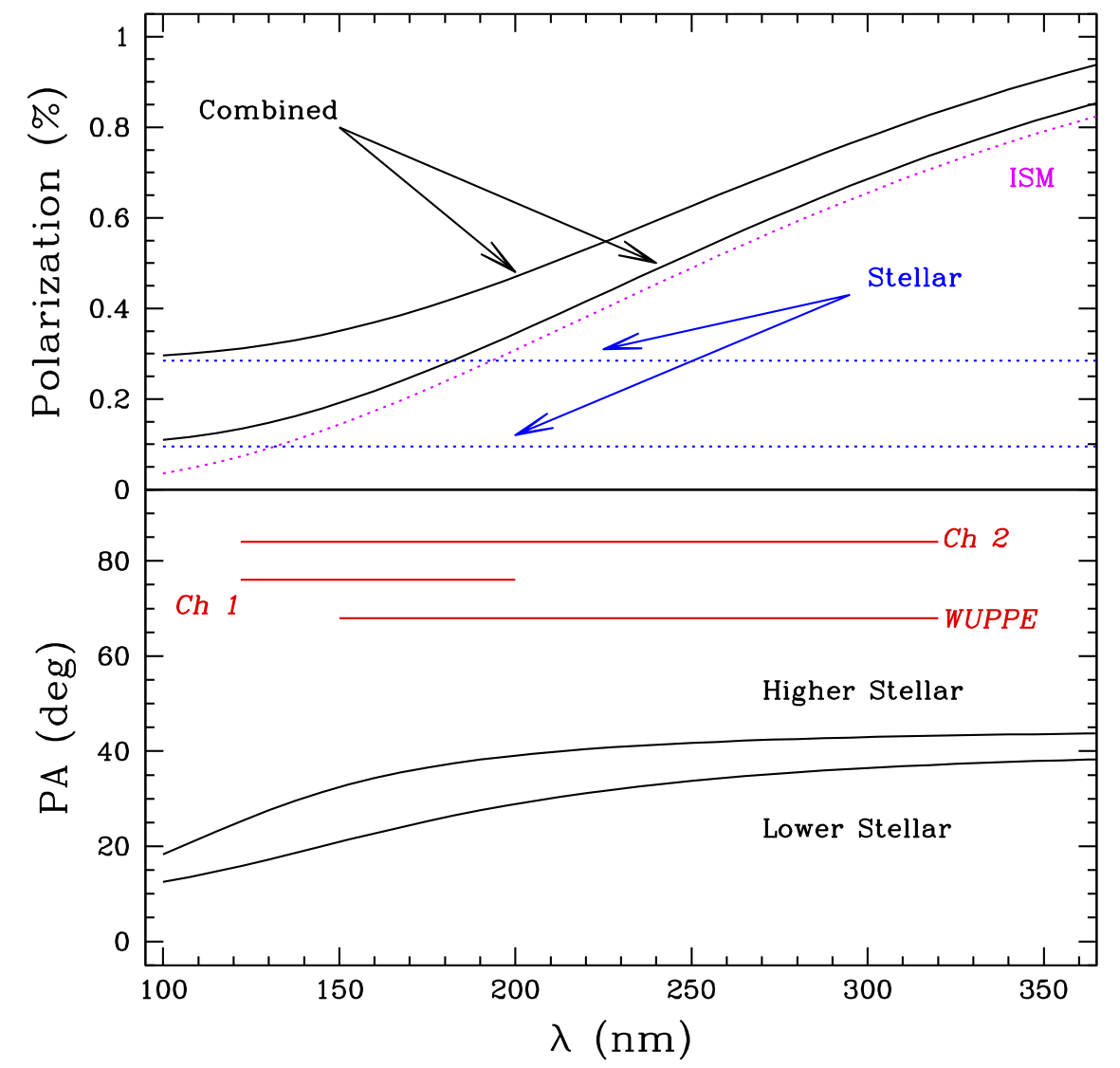}   
\end{center}
  \caption{\small Example of the wavelength dependence of the polarization fraction (top) and the position angle (bottom) for a source with combined ISM and stellar contributions. Red horizontal lines mark the wavelength range of Channels 1 and 2 of Polstar as well as the range of WUPPE. The two contributions can usually be reliably separated based on the spectral shapes of the effects and the different position angles of the components.  For an intrinsic stellar component caused by Thompson scattering (independent of wavelength) overlaid on an interstellar component following the Serkowski form, a rotation of the position angle will occur where the decrease in the latter falls below the constant polarization from the former.  Note that the cross-over wavelength depends on e.g. the ISM extinction value and could occur up into the optical range (Figure courtesy of R. Ignace).}
\label{fig:decomp}
\end{figure}

The addition of optical data from ground-based observatories can extend the leverage of such spectral compositions further, but would, in most cases, not be a requirement.  Figure \ref{fig:decomp} shows a simple model of a combination of stellar (Thompson scattering) and interstellar contributions.
Additional constraints supporting such decompositons can be derived from photometry, imaging, and polarimetry mapping of the target and its immediate surroundings.  As noted already by \citet{hiltner1949b} extinction is a necessary (albeit not sufficient) condition for interstellar polarization.  Thus using modern high precision estimates of the extinction provides significant constraints on the polarization \citep[e.g.][]{Lallement2019A&A...625A.135L, Green2019ApJ...887...93G}. Polarization measurements (especially systematic orientation of the position angles) of the surrounding field stars \citep[e.g.][]{bga1997,nordsieck2001,lomax2017} can be used to set the initial conditions for both the stellar and interstellar polarization position angles in the fitting of the data to Equation \ref{equ:decomp}. 
Several new high sampling surveys of optical polarization are now available or under way \citep[e.g.][]{berdyugin2014,clemens2012b,Southpol,Tassis2018}, 
and large-area imaging polarimeters \citep[e.g. RoboPol;][]{Ramaprakash2019} can be used to acquire dedicated local polarization maps.  Similarly, archival multi-band imaging can be used to indicate and constrain the presence of Rayleigh scattering in and around reflection nebulae \citep{matsumura2011,bga2013}.
Utilizing the high-accuracy stellar distances available from the Gaia mission \citep{GaiaDR2}, the third dimension may also yield constrains on the decomposition.  For instance, if only field stars significantly behind the Polstar target stars show detectable ISM polarization, then the target star polarization spectrum is likely to be of purely stellar (and circumstellar) origin.

\section{Conclusions}\label{sec:concl}

With the availability of an observationally well-supported theory of grain alignment (RAT alignment), interstellar polarimetry has entered a new era of utility.  As we have discussed above, UV polarimetry provides a unique probe of the smallest grains, their mineralogy, dynamics and environment.  Such observations can address the interaction of the small grains with EUV radiation fields, probe the magnetic fields in highly ionized gas, and - finally - establish the carrier(s) of the 2175\AA\ extinction feature.

Since the discovery of interstellar polarization in 1949, the dominant use of such data has been for tracing and measuring magnetic fields \citep{davis1951b,chandrasekhar1953,houde2009}.  This ability remains a strong case for ISM polarimetry.  Because of the wide wavelength range over which dust-induced polarization is observable, multi-wavelength polarimetry provides the possibility of tomographic mapping of the magnetic field, both directly, and through the more indirect use of the parameters of the ``Serkowski curve" \citep{clayton1995,whittet2001,bga2007,vaillancourt2020}.  Here, short-wavelength polarization, especially in the UV, preferentially traces the diffuse and hot ISM, due to low collisional disalignment, low extinction of the radiation field, and the presence of hard aligning radiation (Sec. \ref{sec:SuSeP}).  Exploring the 3D polarization structure \citep[utilizing Gaia distances of the target stars][]{GaiaDR2} of UV observations combined with optical and infrared data will provide important new information on the magnetic field structure of the Galactic ISM.

While we have outlined several problems and experiments that can already be quantitatively formulated and that can be addressed by the Polstar mission, we stress that with a current scarce total sample of 28 lines of sight observed in UV polarimetry, by WUPPE and HST/FOS, a significant discovery space exists for new effects and grain properties, with well defined samples of UV polarimetry probing regions with well characterized environmental parameters.
The Polstar mission will provide radical advances in our understanding of interstellar polarization and the numerous physical effects and parameters probed by it.

\begin{acknowledgements}
B-G A. gratefully acknowledges the support of the National Science Foundation under grant AST-1715876. GVP acknowledges support by NASA through the NASA Hubble Fellowship grant  \#HST-HF2-51444.001-A  awarded  by  the  Space Telescope Science  Institute,  which  is  operated  by  the Association of Universities for Research in Astronomy, Incorporated, under NASA contract NAS5-26555.
AMM's work and Optical/NIR Polarimetry at IAG has been supported over
the years by several grants from S\~ao Paulo state funding agency FAPESP,
especially 01/12589-1 and 10/19694-4. AMM has also been partially supported
by Brazilian agency CNPq (grant 310506/2015-­8).
P.S. acknowledges financial support by the NASA Goddard Space Flight Center to formulate the mission proposal for Polstar.
HY acknowledges partial support by a Templeton senior grant from the Beyond the Horizons program.

\end{acknowledgements}

\bibliographystyle{aasjournal}



\end{document}